\documentclass[11pt,draftcls,onecolumn]{IEEEtran}
\usepackage{amssymb, amsmath, graphicx, cite, color, epsfig}

\newcommand{\complex}{\mathbb{C}}

\newcommand{\Fig}   {\mbox{Fig.} }

\newcommand{\eqdef}{ := }

\newcommand{\thmend}{\hspace*{\fill}~\QEDopen\par\endtrivlist\unskip}
\newcommand{\myproof}[1]{\noindent\hspace{2em}{\itshape #1 }}

\newcommand{\na} { {\sf a} }
\newcommand{\nb} { {\sf b} }
\newcommand{\nr} { {\sf r} }
\newcommand{\prot}{ {\sf P} }
\newcommand{\pa} { {(1)} }
\newcommand{\pb} { {(2)} }
\newcommand{\pc} { {(3)} }
\newcommand{\pd} { {(4)} }

\newtheorem{theorem}{Theorem}
\newtheorem{lemma}[theorem]{Lemma}

\allowdisplaybreaks[1]
\begin{document}
\title{Performance Bounds for Bi-Directional Coded Cooperation Protocols}
\author{Sang Joon Kim,
Patrick Mitran,
and Vahid Tarokh%
\thanks{ Sang Joon Kim and Vahid Tarokh are with the School
of Engineering and Applied Sciences, Harvard University, Cambridge,
MA 02138. Emails:~{sangkim, vahid}@fas.harvard.edu. Patrick Mitran
is with the Department of Electrical and Computer Engineering,
University of Waterloo, Waterloo, Canada.
Email:~pmitran@uwaterloo.ca. This research is supported in part by
NSF grant number ACI-0330244 and ARO MURI grant number
W911NF-07-1-0376. This work was supported in part by the Army
Research Office,under the MURI award N0. N00014-01-1-0859. The views
expressed in this paper are those of the authors alone and not of
the sponsor.} }

%\markboth{IEEE TRANSACTIONS ON INFORMATION THEORY, VOL. XX, NO. YY,
%2007}{Kim, Mitran and Tarokh: Capacity Bounds for Bi-Directional Coded Cooperation}

\maketitle

\begin{abstract}

In coded bi-directional cooperation, two nodes wish to exchange
messages over a shared half-duplex channel with the help of a relay.
In this paper, we derive performance bounds for this problem for each of
three decode-and-forward protocols.

The first protocol is a two phase protocol
where both users simultaneously transmit during the first phase and the relay alone transmits
during the second. In this protocol, our bounds are tight.

The second protocol considers sequential transmissions from the two users followed by a transmission
from the relay while the third protocol is a hybrid of the first two protocols and has four phases.
In the latter two protocols the %inner and outer
bounds are not identical. Numerical evaluation
shows that %at least
in some cases of interest our bounds do not differ significantly.

Finally, in the Gaussian case with path loss, we derive achievable rates and compare
the relative merits of each protocol. This case is of interest in cellular systems.
Surprisingly, we find that in some cases, the achievable rate
region of the four phase protocol contains points that are outside the
outer bounds of the other two protocols.

\end{abstract}


\section{Introduction}
Consider two users, denoted by $\na$ and $\nb$, who wish to share
independent messages over a shared channel. Traditionally, this
problem is known as the two-way channel \cite{cover:2006,
shannon:1961}.
%with well known bounds on its capacity region going
%as far back as \cite{shannon:1961}.
%In \cite{Rankov:2006}, achievable data regions for relay channel are considered with several cooperation strategies %in the two-way communication.

In many realistic broadcast environments, such as wireless communications, it is not unreasonable to
assume the presence of a third node which may aid in the exchange of $\na$ and $\nb$'s messages.
In particular, if $\na$ is a mobile user and $\nb$ is a base station, then we may suppose the
presence of a relay station $\nr$ to assist in the bi-directional communication.

\begin{figure}[t]
\begin{center}
\epsfig{keepaspectratio = true, width = 2in, figure = 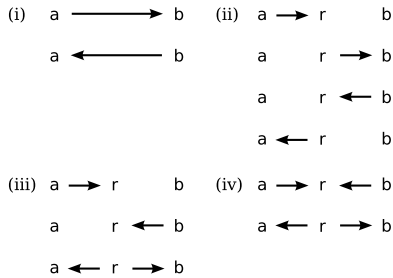}
\end{center}
\caption{(i) Traditional approach, (ii) Naive four phase bi-directional cooperation,
(iii) coded broadcast three phase protocol, (iv) two phase protocol. }
\label{fig:all_protocols}
\end{figure}

Traditionally, without the presence of the relay station, communication between nodes $\na$
and $\nb$ is performed in two steps: first $\na$ transmits its message to $\nb$ followed by
similar transmission from $\nb$ to $\na$ (illustrated in \Fig \ref{fig:all_protocols}.i).
In the presence of relay node $\nr$, one might initially assume that four phases are needed
(see \Fig \ref{fig:all_protocols}.ii). However, by taking advantage of the shared wireless medium,
it is known that the third and fourth transmissions may be combined (\Fig \ref{fig:all_protocols}.iii)
into a single transmission using, for example, ideas from network coding \cite{Ahlswede:2000}, \cite{Xie:2007}.
In particular, if the messages of $\na$ and $\nb$ are $w_\na$ and $w_\nb$ respectively and
belong to a group, then it is sufficient for the relay node to successfully
transmit $w_\na \oplus w_\nb$ simultaneously to $\na$ and $\nb$. In \cite{larsson:2005,larsson:2006},
such a three phase coded bi-directional protocol is considered when the group
is ${\mathbb Z}_2^k$, the binary operator is component-wise modulo 2 addition
(i.e., exclusive or)
and encoding is performed linearly to produce parity bits. As each user transmits sequentially,
each user is amenable to receive ``side-information'' from the opposite user during one of
the first two phases.
%The work in \cite{larsson:2005,larsson:2006}
%then considers methods of combining the soft information thus obtained on the parity bits with that
%from the relay's transmission using methods such as Chase Combining (CC) and show that
%with optimal power allocation, throughput can be increased by up to 33\% over a four phase protocol.

The works of \cite{Popovski:2006b} and \cite{Popovski:2006a} not
only consider the three phase protocol, but combine the first two
phases into a single joint transmission by nodes $\na$ and $\nb$
followed by a single transmission by the relay which forwards its
received signal (\Fig \ref{fig:all_protocols}.iv).
%The forwarding may consist of amplification
%\cite{Popovski:2006a} or symbol by symbol denoising \cite{Popovski:2006b}.
%Further analysis and comparison of the performance of these 3 protocols has
%been carried out in \cite{Popovski:2006b} and \cite{Popovski:2006a}.
Coded bi-directional cooperation may also be extended for the case
of multiple relaying nodes \cite{wu:2004,wu:2005}. In
\cite{Rankov:2006}, achievable rate regions are derived assuming
full duplex capabilities at all nodes.

In this paper, we are interested in determining fundamental bounds
on the performance of coded bi-directional communications assuming
various {\em decode-and-forward} protocols for {\em half-duplex}
channels. In the case of a two phase protocol where both users
transmit simultaneously in the first phase followed by a
transmission from the relay, we derive the exact
performance\footnotemark. In the case of three or more phase
protocols, we take into account any side information that a node may
acquire when it is not transmitting and derive inner and outer
bounds on the capacity regions. We find that a four phase hybrid
protocol is sometimes strictly better than the outerbounds of two or
three phase decode-and-forward protocols previously introduced in
the literature. \footnotetext{
%Recently, we have learned that
Similar results were independently derived in
\cite{Oechtering:2007}. } This paper is structured as follows. In
Section \ref{sec:prelim}, we define our notation and the protocols
that we consider. In Section \ref{sec:bounds}, we derive performance
bounds for the protocols while in Section \ref{sec:gaussian}, we
numerically compute these bounds for fading Gaussian channels.
%Finally, in Section \ref{sec:conclusion}, we summarize our results.

\section{Preliminaries}
\label{sec:prelim}

%In this section, we first introduce our notation and formally define the protocols
%that will be used in the rest of the paper.

\subsection{Notation and Definitions}
We first start with a somewhat more general formulation of the problem. We
consider an $m$ node set, denoted as ${\cal M} \eqdef \{1, 2, \ldots, m\}$ (where $\eqdef$ means defined as) for now, where node $i$
has message $W_{i,j}$ that it wishes to send to node $j$. Each node $i$ has channel input alphabet
${\cal X}^*_i = {\cal X}_i \cup \{ \varnothing \}$ and channel output
alphabet ${\cal Y}^*_i = {\cal Y}_i \cup \{ \varnothing \}$,
where $\varnothing$ is a special symbol distinct of those in ${\cal X}_i$ and ${\cal Y}_i$
and which denotes either no input or no output. In this paper, we assume that a node
may not simultaneously transmit and receive at the same time.
In particular, if node $i$ selects $X_i = \varnothing$, then it receives
$Y_i \in {\cal Y}_i$ and if $X_i \in {\cal X}_i$, then necessarily
$Y_i = \varnothing$, i.e., $X_i = \varnothing$ iff $Y_i \neq \varnothing$\footnotemark.
Otherwise, the effect of one node remaining silent on the received
variable at another node may be arbitrary at this point.
The channel is assumed discrete {\em memoryless}.
In Section IV, we will be interested
in the case ${\cal X}^*_i = {\cal Y}^*_i = \complex \cup \{\varnothing\}$, $\forall i\in {\cal M}$.

\footnotetext{
%Frequency division multiplexing
Thus, FDM cannot be allowed as it violates the half-duplex
constraint. }

The objective of this paper is to determine achievable data rates
and outer bounds on these for some particular cases. We use $R_{i,j}$ for the
transmitted data rate of node $i$ to node $j$, i.e.,
$W_{i,j} \in \{0, \ldots, \lfloor 2^{nR_{i,j}} \rfloor - 1\} \eqdef {\cal S}_{i,j}$.

For a given protocol $\prot$, we denote by $\Delta_\ell \geq 0$ the relative time duration of
the $\ell^{th}$ phase. Clearly, $\sum_\ell \Delta_\ell = 1$. It is also convenient to denote
the transmission at time $k$, $1 \leq k \leq n$ at node $i$ by $X_i^k$, where the total
duration of the protocol is $n$ and $X_i^{(\ell)}$ denotes
the random variable with alphabet ${\cal X}_i^*$ and input distribution $p^{(\ell)}(x_i)$ during phase $\ell$. Also,  $X_i^k$ corresponds to a transmission in the first phase if $k \leq \Delta_1 n$, etc. We also define
$X_{S}^k \eqdef \{X_i^k | i\in S\}$, the set of transmissions by all nodes in the set $S$ at time $k$ and similarly $X_{S}^{(\ell)} \eqdef \{X_i^{(\ell)}|i\in S\}$, a set of random variables with channel input distribution $p^{(\ell)}(x_{S})$ for phase $\ell$, where $x_{S} \eqdef \{x_i | i\in S\}$. Lower case letters $x_i$ denote instances of the upper case $X_i$ which lie in the calligraphic alphabets ${\cal X}_i^*$. Boldface ${\bf x}_i$ represents a vector indexed by time at node $i$. Finally, it is convenient to denote by ${\bf x}_S \eqdef \{{\bf x}_i | i\in S\}$, a set of vectors indexed by time.

Encoders are then given by functions $X_i^k(W_{i,1}, \ldots, W_{i,m}, Y_i^1,\ldots, Y_i^{k-1})$,
for $k = 1, \ldots, n$ and decoders by $\hat{W}_{j,i}(Y_i^1,\ldots, Y_i^{n}, W_{i,1}, \ldots, W_{i,m})$.
Given a block size $n$, a set of encoders and decoders has associated error events
$E_{i,j} \eqdef \{W_{i,j} \neq \hat{W}_{i,j}(\cdot)\}$, for decoding the message $W_{i,j}$
at node $j$ at the end of the block, and the corresponding encoders/decoders result
in relative phase durations $\{\Delta_{\ell, n}\}$, where the subscript $n$ indicates that the phase duration depends on the choice of block size (as they must be multiples of $1/n$).

A set of rates $\{ R_{i,j} \}$ is said to be achievable for a
protocol with phase durations $\{\Delta_\ell\}$, if there exist
encoders/decoders of block length $n=1,2,\ldots$ with $P[E_{i,j}]
\rightarrow 0$ and $\Delta_{\ell,n} \rightarrow \Delta_\ell$ as $n
\rightarrow \infty$  $\forall \ell$. An achievable rate region
(resp. capacity region) is the closure of a set of (resp. all)
achievable rate tuples for fixed $\{\Delta_{\ell}\}$.

\subsection{Basic Results}
In the next section, we will use a variation of the cut-set bound. We assume that all messages from different sources are independent, i.e., $\forall i\neq j$, $W_{i,k}$ and $W_{j,l}$ are independent $\forall k,l \in {\cal M}$. In contrast to \cite{cover:2006}, we relax the independent assumption from one source to different nodes, i.e., in our case $W_{i,j}$ and $W_{i,k}$ may not be independent. Given subsets $S,T \subseteq {\cal M}$, we define $W_{S,T} \eqdef \{W_{i,j} | i\in S , j\in T\}$ and $R_{S,T} = \lim_{n\rightarrow \infty }\frac1n H(W_{S,T})$.

\begin{lemma}
\label{lemma:cover:1}
If in some network the information rates $\{R_{i,j}\}$ are achievable for a protocol $\prot$
with relative durations $\{\Delta_\ell\}$, then for every $\epsilon>0$ and all $S \subset \{1,2,\cdots,m\} = {\cal M}$
\begin{align}
 R_{S,S^c} \leq
   \sum_\ell \Delta_\ell I(X^{(\ell)}_{S};Y^{(\ell)}_{S^c}|X^{(\ell)}_{S^c},Q)+\epsilon,
\end{align}
for a family of conditional distributions $p^{(\ell)}(x_1, x_2, \ldots, x_m|q)$
and a discrete time-sharing random variable $Q$ with distribution $p(q)$.
Furthermore, each $p^{(\ell)}(x_1, x_2, \ldots, x_m|q)p(q)$ must satisfy the
constraints of phase $\ell$ of protocol $\prot$.
\thmend
\end{lemma}
\begin{proof}
Replacing $W^{(T)}$ by $W_{S,S^c}$ and $W^{(T^c)}$ by $W_{S^c,{\cal M}}$
in (15.323) - (15.332) in \cite{cover:2006}, then all the steps in \cite{cover:2006} still hold and we have
\begin{align*}
H(W_{S,S^c}) = H(W_{S,S^c}|W_{S^c,{\cal M}}) \leq \sum_{k=1}^{n} I(X_{S}^k;Y_{S^c}^k|X_{S^c}^k) + n\epsilon_n,
\end{align*}
where $\epsilon_n \rightarrow 0$ since $\sum_{i \in S, j\in S^c} P[E_{i,j}] \rightarrow 0$ and
the distributions $p(x_1^k, \ldots, x_m^k, y_1^k, \ldots, y_m^k)$ are those induced
by encoders for which $P[E_{i,j}] \rightarrow 0$ as $n \rightarrow \infty$.

Defining $Q_1, Q_2, \ldots$ to be discrete random variables
uniform over $\{1, \ldots ,n\cdot\Delta_{1,n}\}, \{n\cdot\Delta_{1,n}+1,\ldots,n\cdot \Delta_{1,n}+n\cdot\Delta_{2,n}\}, \ldots$,
we thus have
\begin{align}
H(W_{S,S^c}) \leq
    \sum_\ell n\cdot\Delta_{\ell,n} I(X_{S}^{Q_\ell};Y_{S^c}^{Q_\ell}|X_{S^c}^{Q_\ell},Q_\ell) + n\epsilon_n,
\end{align}
Defining the discrete random variable $Q \eqdef (Q_1, Q_2, \ldots)$,
then
\begin{align}
\frac{1}{n}H(W_{S,S^c}) \leq
    \sum_\ell \Delta_{\ell,n} I(X_{S}^{(\ell)};Y_{S^c}^{(\ell)}|X_{S^c}^{(\ell)}, Q) + \epsilon_n,
\end{align}
where $X_{S}^{(\ell)} \eqdef X_{S}^{Q_\ell}$.
Finally, since the distributions  $p^{(\ell)}(x_1, x_2, \ldots, x_m|q)p(q)$ are
those induced by encoders for which $P[E_{i,j}] \rightarrow 0$, if there is
a constraint on the encoders (such as a power constraint), this constraint is also
valid for the distributions $p^{(\ell)}(x_1, x_2, \ldots, x_m|q)p(q)$.
\end{proof}

% We will also require the following well-known theorem.
% \begin{theorem}\label{lemma:cover:2}
% $(Carath\acute{e}odory)$: Any point in the convex closure of a
% connected compact set $A$ in a $d$ dimensional Euclidean space can
% be represented as a convex combination of $d+1$ or fewer points in
% the original set $A$.
% \end{theorem}

\subsection{Protocols}
In bi-directional cooperation, two terminal nodes denoted $\na$ and $\nb$ exchange their messages. The messages to be transmitted are $W_\na \eqdef W_{\na,\nb}$, $W_\nb \eqdef W_{\nb,\na}$ and the corresponding rates are $R_\na \eqdef R_{\na,\nb}$ and $R_\nb \eqdef R_{\nb,\na}$. The two distinct messages $W_\na$ and
$W_\nb$ are taken to be independent and uniformly distributed in the set of $\{0,\ldots,\lfloor 2^{nR_\na}\rfloor  -1\}\eqdef {\cal S}_\na$ and $\{0,\ldots,\lfloor 2^{nR_\nb} \rfloor -1\}\eqdef {\cal S}_\nb$, respectively. Then  $W_\na$ and $W_\nb$ are both members of the additive group ${\mathbb Z}_L$, where
$L = \max(\lfloor 2^{nR_\na} \rfloor, \lfloor 2^{nR_\nb} \rfloor)$.

\begin{figure}[t]
\begin{center}
\epsfig{keepaspectratio = true, width = 4 in, figure =
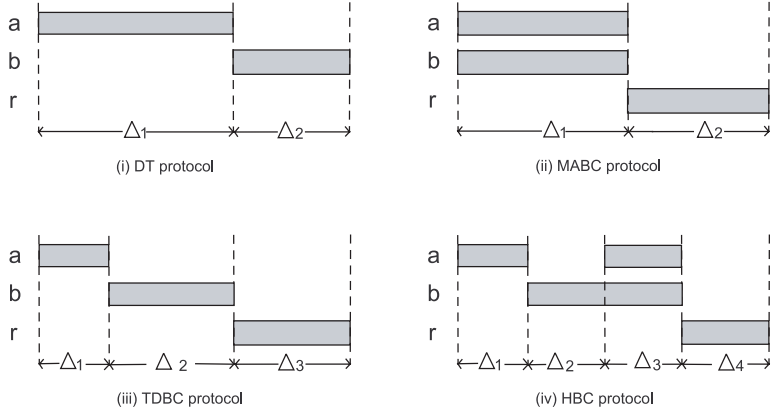}
\end{center}
\caption{Proposed protocol diagrams. Shaded areas denote transmission
by the respective nodes. It is assumed that all nodes listen when not
transmitting.}
\label{fig:BC_protocols}
\end{figure}

The simplest protocol for the bi-directional channel, is that of
Direct Transmission (DT) (\Fig \ref{fig:BC_protocols}.i). Here,
since the channel is memoryless and $\epsilon > 0$ is arbitrary, the
capacity region from Lemma \ref{lemma:cover:1} is :
\begin{align*}
R_\na &\leq \sup_{p^\pa(x_\na)} \Delta_1 I(X_\na^\pa;Y_\nb^\pa|X_\nb^\pa = \varnothing)\\
R_\nb &\leq \sup_{p^\pb(x_\nb)} \Delta_2
I(X_\nb^\pb;Y_\na^\pb|X_\na^\pb = \varnothing),
\end{align*}
where the distributions are over the alphabets ${\cal X}_\na$ and
${\cal X}_\nb$ respectively.

With a relay node $\nr$, we suggest three different
decode-and-forward protocols, which we denote as Multiple Access
Broadcast (MABC) protocol, Time Division Broadcast (TDBC) and Hybrid
Broadcast (HBC). Then, the message from $\na$ (resp. $\nb$) to $\nr$
is $W_{\na,\nr} = W_\na$ (resp. $W_{\nb,\nr} = W_\nb$) and the
corresponding rate is $R_{\na,\nr} = R_\na$ (resp. $R_{\nb,\nr} =
R_\nb$). Also, in our protocols, all phases are contiguous, i.e.,
they are performed consecutively and are not interleaved or
re-ordered.\footnote {If we relax the contiguous assumption, the
achievable region could increase by cooperation between interleaving
phases.}

In the MABC protocol (\Fig \ref{fig:BC_protocols}.ii), terminal nodes $\na$
and $\nb$ transmit information simultaneously during phase 1 and the
relay $\nr$ transmits some function of the received signals during phase 2. With
this scheme, we only divide the total time period into two regimes and
neither node $\na$ nor node $\nb$ is able to receive
any meaningful side-information during the first phase due to the half-duplex
constraint.

In the TDBC protocol (\Fig \ref{fig:BC_protocols}.iii), only node
$\na$ transmits during the first phase and only node $\nb$ transmits
during the second phase. In phase 3, relay $\nr$ performs a
transmission based on the received data from the first two phases.
Here, node $\na$ attempts to recover the message $W_\nb$ based on
both the transmissions from node $\nb$ in the second phase and node
$\nr$ in the third phase. We denote the received signal at node
$\na$ in the second phase as second phase side information.
Likewise, node $\nb$ may also recover $W_\na$ based on first phase
side information and the received signal at node $\nb$ during the
third phase.

Finally, we consider a Hybrid Broadcast (HBC) protocol (\Fig \ref{fig:BC_protocols}.iv)
which is an amalgam of the MABC and TDBC protocols. In this scheme, there are
4 distinct transmissions, two of which result in side-information at $\na$ and $\nb$.

\section{Performance Bounds}
\label{sec:bounds}

% In this section, we derive an achievable region and outer bound for each protocol.

\subsection{MABC Protocol}

\begin{theorem}
\label{theorem:MABC}
The capacity region of the half-duplex bi-directional relay channel
with the MABC protocol is the closure of the set of all points $(R_\na,R_\nb)$ satisfying
\begin{align*}
R_{\na} &< \min \big\{ \Delta_1 I(X_{\na}^\pa ; Y_{\nr}^\pa |
X_{\nb}^\pa,X_\nr^\pa = \varnothing,Q), \Delta_2 I(X_{\nr}^\pb ; Y_{\nb}^\pb |X_\na^\pb = X_\nb^\pb = \varnothing, Q)\big\} \\
R_{\nb} &< \min \big\{ \Delta_1 I(X_{\nb}^\pa ; Y_{\nr}^\pa |
X_{\na}^\pa,X_\nr^\pa = \varnothing,Q),\Delta_2 I(X_{\nr}^\pb ; Y_{\na}^\pb |X_\na^\pb = X_\nb^\pb = \varnothing, Q)\big\}\\
R_{\na} + R_{\nb} &<    \Delta_1 I(X_{\na}^\pa , X_{\nb}^\pa ; Y_{\nr}^\pa|X_\nr^\pa = \varnothing,Q)
\end{align*}
over all joint distributions
$p(q)p^\pa(x_{\na}|q)p^\pa(x_{\nb}|q)p^\pb(x_{\nr}|q)$ with $|{\cal
Q}| \leq 5$ over the alphabet ${\cal X}_\na \times {\cal X}_\nb
\times {\cal X}_\nr $. \thmend
\end{theorem}

{\em Remark:}
If the relay is not required to decode both messages, then the region above is still achievable, and removing the constraint on the sum-rate $R_\na + R_\nb$ yields an outer bound.
%An example of when this is tight for a situation of theoretical interest is provided in Appendix \ref{app:MABC}

\begin{proof}
{\em Achievability:} {\em Random code generation: } For simplicity
of exposition only, we take $|{\cal Q}| = 1$ and therefore consider
distributions $p^\pa(x_\na)$, $p^\pa(x_\nb)$ and $p^\pb(x_\nr)$.
First we generate random $(n\cdot \Delta_{1,n})$-length sequences
$\mathbf{x}^{(1)}_\na(w_\na)$ with $w_\na \in S_\na$ and
$\mathbf{x}^{(1)}_\nb(w_\nb)$ with $w_\nb \in S_\nb$, and
$(n\cdot\Delta_{2,n})$-length sequences
$\mathbf{x}^{(2)}_\nr(w_\nr)$ with $w_\nr \in \mathbb{Z}_L$ where $L
= \max(\lfloor 2^{nR_{\na}} \rfloor, \lfloor 2^{nR_{\nb}} \rfloor
)$, according to $p^\pa(x_\na)$, $p^\pa(x_\nb)$ and $p^\pb(x_\nr)$
respectively.

{\em Encoding: } During phase 1, encoders of node $\na$ and $\nb$ send the codewords $\mathbf{x}^{(1)}_\na(w_\na)$ and $\mathbf{x}^{(1)}_\nb(w_\nb)$ respectively. Relay $\nr$ estimates $\hat{w}_\na$ and $\hat{w}_\nb$ after phase 1 using jointly typical decoding, then constructs $w_{\nr} =\hat{w}_\na \oplus \hat{w}_\nb$ in ${\mathbb Z}_L$ and sends
$\mathbf{x}^{(2)}_\nr(w_\nr)$ during phase 2.

{\em Decoding: } $\na$ and $\nb$ estimate $\tilde{w}_\nb$ and $\tilde{w}_\na$ after phase 2 using jointly typical decoding.
Since $w_\nr = w_\na \oplus w_\nb$ and $\na$ knows $w_\na$, node $\na$ can reduce the number of possible
$w_\nr$ to $\lfloor 2^{nR_{\nb}} \rfloor$ and likewise at node $\nb$, the cardinality is $\lfloor 2^{nR_{\na}} \rfloor$.

{\em Error analysis: }  For convenience of analysis, first define
$E_{i,j}^{(\ell)}$ as the error event at node $j$ that node $j$
attempts to decode $w_i$ at the end of phase $\ell$ using jointly
typical decoding. Let $A^{(\ell)}_{S,T}$ represents the set of
$\epsilon$-weakly typical
$(\mathbf{x}_{S}^{(\ell)},\mathbf{y}_{T}^{(\ell)})$ sequences of
length $n \cdot \Delta_{\ell,n}$ according to the input
distributions employed in phase $\ell$. Also define the set of
codewords ${\bf x}_S^{(\ell)}(w_S) := \{ {\bf x}_i^{(\ell)}(w_i) | i
\in S \}$ and the events $D_{S,T}^{(\ell)}(w_{S}) \eqdef \{({\bf
x}^{(\ell)}_{S}(w_{S}),\mathbf{y}^{(\ell)}_{T}) \in
A_{S,T}^{(\ell)}\}$, where $S$ and $T$ are disjoint subsets of
nodes.
\begin{align}
  P[E_{\na,\nb}] & \leq P[E_{\na,\nr}^\pa \cup E_{\nb,\nr}^\pa \cup E_{\nr,\nb}^\pb]\\
  & \leq P[E_{\na,\nr}^\pa \cup E_{\nb,\nr}^\pa] + P[E_{\nr,\nb}^\pb | \bar{E}_{\na,\nr}^\pa \cap \bar{E}_{\nb,\nr}^\pa]
 % P[E_{\nb,\na}]& \leq P[E_{\na,\nr}^\pa \cup E_{\nb,\nr}^\pa \cup E_{\nr,\na}^\pb]\\
 % & \leq P[E_{\na,\nr}^\pa \cup E_{\nb,\nr}^\pa] + P[E_{\nr,\na}^\pb | \bar{E}_{\na,\nr}^\pa \cap \bar{E}_{\nb,\nr}^\pa]
\end{align}
Following the well-known MAC error analysis from (15.72) in \cite{cover:2006}:
\begin{align}
  P[E_{\na,\nr}^\pa \cup E_{\nb,\nr}^\pa] \leq & P[\bar{D}^\pa_{\{\na,\nb\},\{\nr\}}(w_\na,w_\nb)] + 2^{nR_\na} 2^{-n\cdot\Delta_{1,n}(I(X_\na^\pa;Y_\nr^\pa|X_\nb^\pa,X_\nr^\pa = \varnothing)-3\epsilon)} +\nonumber\\
   &2^{nR_\nb} 2^{-n\cdot\Delta_{1,n}(I(X_\nb^\pa;Y_\nr^\pa|X_\na^\pa,X_\nr^\pa = \varnothing)-3\epsilon)} +  2^{n(R_\na+R_\nb)} 2^{-n\cdot\Delta_{1,n}(I(X_\na^\pa,X_\nb^\pa;Y_\nr^\pa|X_\nr^\pa =
   \varnothing)-4\epsilon)}\label{eq:union0}
\end{align}
  Also,
\begin{align}
 % P[&E_{\nr,\na}^\pb | \bar{E}_{\na,\nr}^\pa \cap \bar{E}_{\nb,\nr}^\pa] \nonumber\\
 % &\leq P[\bar{D}^\pb_{\{\nr\},\{\na\}}(w_\na \oplus w_\nb)] + P[\cup_{\tilde{w}_\nb \neq w_\nb} D_{\{\nr\},\{\na\}}^\pb(w_\na \oplus \tilde{w}_\nb)] \nonumber\\
 % &\leq P[\bar{D}^\pb_{\{r\},\{\na\}}(w_\na \oplus w_\nb)] + \nonumber\\
 % &~~~~2^{nR_\nb} 2^{-n\cdot\Delta_{2,n} (I(X_\nr^\pb;Y_\na^\pb|X_\na^\pb = X_\nb^\pb = \varnothing)-3\epsilon)} \label{eq:union1}
  P[E_{\nr,\nb}^\pb | \bar{E}_{\na,\nr}^\pa \cap \bar{E}_{\nb,\nr}^\pa] &\leq P[\bar{D}^\pb_{\{\nr\},\{\nb\}}(w_\na \oplus w_\nb)] + P[\cup_{\tilde{w}_\na \not= w_\na} D_{\{\nr\},\{\nb\}}^\pb(\tilde{w}_\na \oplus w_\nb)] \nonumber\\
  &\leq P[\bar{D}^\pb_{\{\nr\},\{\nb\}}(w_\na \oplus w_\nb)] + 2^{nR_\na} 2^{-n\cdot\Delta_{2,n}(I(X_\nr^\pb;Y_\nb^\pb|X_\na^\pb = X_\nb^\pb = \varnothing)-3\epsilon)} \label{eq:union1}
\end{align}
  Since $\epsilon > 0$ is arbitrary, with the conditions of Theorem \ref{theorem:MABC} and the AEP property, we can make
the right hand sides of \eqref{eq:union0} -- \eqref{eq:union1} tend
to 0 as $n \rightarrow \infty$. Similarly, $P[E_{\nb,\na}]
\rightarrow 0$ as $n \rightarrow \infty$.
%$P[E_{\na,\nr}^\pa \cup E_{\nb,\nr}^\pa]$, $P[E_{\nr,\na}^\pb | \bar{E}_{\na,\nr}^\pa \cap \bar{E}_{\nb,\nr}^\pa]$ and
%$P[E_{\nr,\nb}^\pb | \bar{E}_{\na,\nr}^\pa \cap \bar{E}_{\nb,\nr}^\pa]$ tend to 0 as $n \rightarrow \infty$.

{\em Converse:} We use Lemma~\ref{lemma:cover:1} to prove the
converse part of Theorem~\ref{theorem:MABC}. As we have 3 nodes,
there are 6 cut-sets, $S_1 = \{\na\}$, $S_2 = \{\nb\}$, $S_3 =
\{\nr\}$, $S_4 = \{\na,\nb\}$, $S_5 = \{\na,\nr\}$ and $S_6 =
\{\nb,\nr\}$, as well as two rates $R_\na$ and $R_\nb$. The outer
bound corresponding to $S_1$ is then
\begin{align}
R_\na   &\leq \Delta_1 I(X_\na^\pa; Y_\nr^\pa, Y_\nb^\pa|X_\nr^\pa, X_\nb^\pa, Q) + \Delta_2 I(X_\na^\pb; Y_\nr^\pb, Y_\nb^\pb|X_\nr^\pb, X_\nb^\pb, Q)+\epsilon \\
    &= \Delta_1 I(X_\na^\pa; Y_\nr^\pa | X_\nb^\pa,X_\nr^\pa = \varnothing ,Q)+\epsilon, \label{MABC:3}
\end{align}
where \eqref{MABC:3} follows since in the MABC protocol, we must
have
\begin{align}
 Y_\na^\pa &= Y_\nb^\pa = X_\nr^\pa = \varnothing \\
 X_\na^\pb &= X_\nb^\pb = Y_\nr^\pb = \varnothing. \label{MABC:5}
\end{align}
% [[ I changed this a little so now it fits: ]]
We find the outer bounds of the other cut-sets in the same manner:
\begin{align}
S_2 &: R_\nb \leq \Delta_1 I(X_\nb^\pa;Y_\nr^\pa|X_\na^\pa,X_\nr^\pa = \varnothing ,Q)+\epsilon. \label{MABC:6}\\
S_3 &: N/A  \\
S_4 &: R_\na + R_\nb  \leq\Delta_1 I(X_\na^\pa,X_\nb^\pa;Y_\nr^\pa|X_\nr^\pa = \varnothing, Q)+\epsilon, \label{MABC:7}\\
S_5 &: R_\na  \leq \Delta_2 I(X_\nr^\pb;Y_\nb^\pb| X_\na^\pb = X_\nb^\pb = \varnothing,Q)+\epsilon  , \label{MABC:8}\\
S_6 &: R_\nb \leq \Delta_2 I(X_\nr^\pb;Y_\na^\pb|X_\na^\pb =
X_\nb^\pb = \varnothing,Q) +\epsilon. \label{MABC:9}
\end{align}
Since $\epsilon > 0$ is arbitrary, together, \eqref{MABC:3},
\eqref{MABC:6} -- \eqref{MABC:9} and the fact that the half-duplex
nature of the channel constrains $X_\na^\pa$ to be conditionally
independent of $X_\nb^\pa$ given $Q$ yields the converse. By
Fenchel-Bunt's theorem in \cite{Hiriart:2001}, it is sufficient to
restrict $|{\cal Q}| \leq 5$.
\end{proof}
\subsection{TDBC Protocol}
% In this case, our bounds are not tight, hence we have an achievable region and an outer bound.

\begin{theorem}\label{theorem:TDBC:1}
An achievable region of the half-duplex bi-directional relay channel with the TDBC protocol is
the closure of the set of all points $(R_\na,R_\nb)$ satisfying
\begin{align*}
R_\na <\min\big\{&{\Delta}_1 I(X_\na^\pa;Y_\nr^\pa|X_\nb^\pa = X_\nr^\pa = \varnothing,Q),\\
&{\Delta}_1 I(X_\na^\pa;Y_\nb^\pa|X_\nb^\pa = X_\nr^\pa = \varnothing,Q) +
{\Delta}_3 I(X_\nr^\pc;Y_\nb^\pc|X_\na^\pc = X_\nb^\pc = \varnothing,Q)\big\}\\
R_\nb < \min\big\{&{\Delta}_2 I(X_\nb^\pb;Y_\nr^\pb|X_\na^\pb = X_\nr^\pb = \varnothing,Q),\\
&{\Delta}_2 I(X_\nb^\pb;Y_\na^\pb|X_\na^\pb = X_\nr^\pb =
\varnothing,Q) + 
{\Delta}_3 I(X_\nr^\pc;Y_\na^\pc|X_\na^\pc = X_\nb^\pc =
\varnothing,Q)\big\}
\end{align*}
over all joint distributions
$p(q)p^\pa(x_{\na}|q)p^\pb(x_{\nb}|q)p^\pc(x_{\nr}|q)$ with $|{\cal
Q}| \leq 4$ over the alphabet ${\cal X}_\na \times {\cal X}_\nb
\times {\cal X}_\nr $. \thmend
\end{theorem}
\begin{proof}
{\em Random code generation:} First, we generate
a partition of ${\cal S}_\na$ randomly by independently assigning every index $w_\na \in {\cal S}_\na$
to a set ${\cal S}_{\na,i}$, with a uniform distribution over the indices $i \in
\{0, \ldots, \lfloor 2^{nR_{\na0}} \rfloor - 1\}$. We denote by $s_\na(w_\na)$ the index $i$ of
${\cal S}_{\na,i}$ to which $w_\na$ belongs and likewise, a partition for $w_\nb \in {\cal S}_\nb$ is
similarly constructed.
For simplicity of exposition, we take $|{\cal Q}| = 1$. For any $\epsilon > 0$ and
distributions $p^\pa(x_\na)$ , $p^\pb(x_\nb)$ and $p^\pc(x_\nr)$, we generate random
$(n\cdot\Delta_{1,n})$-length sequences $\mathbf{x}^{(1)}_\na(w_\na)$ with $w_\na \in {\cal S}_\na$,
$(n\cdot\Delta_{2,n})$-length sequences $\mathbf{x}^{(2)}_\nb(w_\nb)$ with $w_\nb \in {\cal S}_\nb$ and
$(n\cdot\Delta_{3,n})$-length sequences $\mathbf{x}^{(3)}_\nr(w_\nr)$ with $w_\nr \in {\mathbb Z}_L$,
$L = \lfloor 2^{n\cdot\max \{R_{a0}, R_{b0}\}} \rfloor$.

{\em Encoding:} During phase 1 (resp. phase 2), the encoder at node $\na$ (resp. node $\nb$)
sends the codeword $\mathbf{x}_\na^\pa(w_\na)$ (resp $\mathbf{x}_\nb^\pb(w_\nb)$).
Relay $\nr$ estimates $\hat{w}_\na$ and
$\hat{w}_\nb$ after phases 1 and 2 respectively. The relay then constructs $w_{\nr} =
s_\na(\hat{w}_\na) \oplus s_\nb(\hat{w}_\nb)$ in ${\mathbb
Z}_L$, and sends $\mathbf{x}_\nr^\pc(w_\nr)$ during phase 3.

{\em Decoding:} Terminal nodes $\na$ and $\nb$ estimate the indices
$\tilde{s}_\nb(w_\nb)$ and $\tilde{s}_\na(w_\na)$ after phase 3 from
${\bf x}^{(3)}_\nr$ and then decode $\tilde{w}_\nb$ and
$\tilde{w}_\na$ if there exists a unique $\tilde{w}_\nb \in
S_{\nb,\tilde{s}_\nb} \cap A^\pb_{\{\nb\},\{\na\}}$ and
$\tilde{w}_\na \in S_{\na,\tilde{s}_\na} \cap
A^\pa_{\{\na\},\{\nb\}}$.

{\em Error analysis: } Define $E_{i,j}^{(\ell)}$ as the error events
from node $i$ to node $j$ assuming node $j$ attempts to decode $w_i$
at the end of phase $\ell$ using jointly typical decoding and
${\tilde s}_\na$ or ${\tilde s}_\nb$ if available. Also we use the
same definitions of $A^{(\ell)}_{S,T}$ and $D_{S,T}^{(\ell)}(w_{S})$
as in the proof of Theorem \ref{theorem:MABC}. Then :
\begin{align}
  P[E_{\na,\nb}]  \leq &P[E_{\na,\nr}^\pa \cup E_{\nb,\nr}^\pb \cup E_{\nr,\nb}^\pc \cup E_{\na,\nb}^\pc]\\
   \leq &P[E_{\na,\nr}^\pa] + P[E_{\nb,\nr}^\pb] + P[E_{\nr,\nb}^\pc | \bar{E}_{\na,\nr}^\pa \cap \bar{E}_{\nb,\nr}^\pb] + P[E_{\na,\nb}^\pc | \bar{E}_{\na,\nr}^\pa \cap \bar{E}_{\nb,\nr}^\pb \cap \bar{E}_{\nr,\nb}^\pc].
\end{align}
Also
\begin{align}
  P[E_{\na,\nr}^\pa] &\leq  P[\bar{D}^\pa_{\{\na\},\{\nr\}}(w_\na)] + 2^{nR_\na} 2^{-n\cdot\Delta_{1,n} (I(X_\na^\pa;Y_\nr^\pa|X_\nb^\pa=X_\nr^\pa = \varnothing)-3\epsilon)} \label{eq:tdbc1} \\
  P[E_{\nb,\nr}^\pb] &\leq  P[\bar{D}^\pb_{\{\nb\},\{\nr\}}(w_\nb)] + 2^{nR_\nb} 2^{-n\cdot\Delta_{2,n} (I(X_\nb^\pb;Y_\nr^\pb|X_\na^\pb=X_\nr^\pb =
  \varnothing)-3\epsilon)}\\
  P[E_{\nr,\nb}^\pc | \bar{E}_{\na,\nr}^\pa \cap
  \bar{E}_{\nb,\nr}^\pb] &\leq P[\bar{D}^\pc_{\{\nr\},\{\nb\}}(s_\na(w_\na)\oplus
s_\nb(w_\nb))]
  + P[\cup_{\tilde{s}_\na \neq s_\na(w_\na)} D_{\{\nr\},\{\nb\}}^\pc(\tilde{s}_\na \oplus
  s_\nb(w_\nb))]\nonumber\\
  &\leq P[\bar{D}^\pc_{\{\nr\},\{\nb\}}(s_\na(w_\na)\oplus s_\nb(w_\nb))] + 2^{nR_{\na 0}} 2^{-n\cdot\Delta_{3,n}(I(X_\nr^\pc;Y_\nb^\pc|X_\na^\pc=X_\nb^\pc = \varnothing)-3\epsilon)}\\
   P[E_{\na,\nb}^\pc | \bar{E}_{\na,\nr}^\pa \cap \bar{E}_{\nb,\nr}^\pb \cap \bar{E}_{\nr,\nb}^\pc] 
   &\leq P[\bar{D}^\pa_{\{\na\},\{\nb\}}(w_\na)] + P[\cup_{\tilde{w}_\na \neq w_\na} D_{\{\na\},\{\nb\}}^\pa(\tilde{w}_\na), s_\na(w_\na) = s_\na(\tilde{w}_\na)]\nonumber\\
  % &\leq P[\bar{D}^\pa_{\{\na\},\{\nb\}}(w_\na)] + \nonumber\\
  % &~~~~\lfloor 2^{nR_\na} \rfloor 2^{-n\cdot\Delta_{1,n} (I(X_\na^\pa;Y_\nb^\pa|X_\nb^\pa=X_\nr^\pa = \varnothing) - 3\epsilon)} 2^{-nR_{\na0}}\nonumber\\
   &\leq  P[\bar{D}^\pa_{\{\na\},\{\nb\}}(w_\na)] + 
   2^{n(R_\na - \Delta_{1,n} I(X_\na^\pa;Y_\nb^\pa|X_\nb^\pa=X_\nr^\pa = \varnothing) - R_{\na0} + 3 \epsilon)} \label{eq:tdbc2}
\end{align}
  Since $\epsilon > 0$ is arbitrary, with the proper choice of $R_{\na0}$, the conditions of Theorem \ref{theorem:TDBC:1} and the AEP property, we can make the right hand sides of \eqref{eq:tdbc1} -- \eqref{eq:tdbc2} vanish as $n \rightarrow \infty$.
%$P[E_{\na,\nr}^\pa]$, $P[E_{\nb,\nr}^\pb]$, $P[E_{\nr,\nb}^\pc | \bar{E}_{\na,\nr}^\pa \cap \bar{E}_{\nb,\nr}^\pb]$ and
%$P[E_{\na,\nb}^\pc | \bar{E}_{\na,\nr}^\pa \cap \bar{E}_{\nb,\nr}^\pb \cap \bar{E}_{\nr,\nb}^\pc]$ tend to 0 as $n \rightarrow \infty$.
Similarly, $P[E_{\nb,\na}] \rightarrow 0$ as $n \rightarrow \infty$.
By Fenchel-Bunt's theorem in \cite{Hiriart:2001}, it is sufficient
to restrict $|{\cal Q}| \leq 4$.
\end{proof}
\begin{theorem}
\label{theorem:TDBC:2} The capacity region of the bi-directional
relay channel with the TDBC protocol is outer bounded by the union
of
\begin{align*}
R_\na \leq \min\{ &{\Delta}_1 I(X_\na^\pa;Y_\nr^\pa,Y_\nb^\pa|X_\nb^\pa = X_\nr^\pa = \varnothing,Q),\\
          & {\Delta}_1 I(X_\na^\pa;Y_\nb^\pa|X_\nb^\pa = X_\nr^\pa = \varnothing,Q) + 
          {\Delta}_3 I(X_\nr^\pc;Y_\nb^\pc|X_\na^\pc = X_\nb^\pc = \varnothing,Q) \} \\
R_\nb \leq \min\{ &{\Delta}_2 I(X_\nb^\pb;Y_\nr^\pb,Y_\na^\pb|X_\na^\pb = X_\nr^\pb = \varnothing,Q),\\
           &{\Delta}_2 I(X_\nb^\pb;Y_\na^\pb|X_\na^\pb = X_\nr^\pb = \varnothing,Q) + 
           {\Delta}_3 I(X_\nr^\pc;Y_\na^\pc|X_\na^\pc = X_\nb^\pc = \varnothing,Q)\} \\
R_\na + R_\nb \leq &\Delta_1 I(X_\na^\pa; Y_\nr^\pa |X_\nb^\pa =
X_\nr^\pa = \varnothing, Q) + \Delta_2 I(X_\nb^\pb; Y_\nr^\pb|X_\na^\pb = X_\nr^\pb =
\varnothing, Q)
\end{align*}
over all joint distributions $p(q)p^\pa(x_{\na}|q)p^\pb(x_{\nb}|q)$
$p^\pc(x_{\nr}|q)$ with $|{\cal Q}| \leq 5$ over the alphabet ${\cal
X}_\na \times {\cal X}_\nb \times {\cal X}_\nr$. \thmend
\end{theorem}
{\em Remark:} If the relay is not required to decode both messages,
removing the constraint on the sum-rate $R_\na + R_\nb$ yields an
outer bound.

\myproof{Proof outline:} The proof of Theorem \ref{theorem:TDBC:2}
follows the same argument as in the proof of the converse part of
Theorem \ref{theorem:MABC}.
\endproof

\subsection{HBC Protocol}
\begin{theorem}\label{theorem:HBC:1}
An achievable region of the half-duplex bi-directional relay channel
with the HBC protocol is the closure of the set of all points $(R_\na,R_\nb)$ satisfying
\begin{align*}
R_\na < \min \big\{& \Delta_1 I(X_\na^\pa;Y_\nr^\pa|X_\nb^\pa = X_\nr^\pa = \varnothing,Q) + \Delta_3 I(X_\na^\pc;Y_\nr^\pc|X_\nb^\pc,X_\nr^\pc = \varnothing,Q), \\
             &\Delta_1 I(X_\na^\pa;Y_\nb^\pa|X_\nb^\pa = X_\nr^\pa = \varnothing,Q) +\Delta_4 I(X_\nr^\pd;Y_\nb^\pd|X_\na^\pd = X_\nb^\pd = \varnothing,Q)\big\} \\
R_\nb < \min \big\{ & \Delta_2 I(X_\nb^\pb;Y_\nr^\pb|X_\na^\pb = X_\nr^\pb = \varnothing,Q) +\Delta_3 I(X_\nb^\pc;Y_\nr^\pc|X_\na^\pc,X_\nr^\pc = \varnothing,Q),\\
             &\Delta_2 I(X_\nb^\pb;Y_\na^\pb|X_\na^\pb = X_\nr^\pb = \varnothing,Q) + \Delta_4 I(X_\nr^\pd;Y_\na^\pd|X_\na^\pd = X_\nb^\pd = \varnothing,Q)\big\} \\
R_\na+R_\nb <&\Delta_1 I(X_\na^\pa;Y_\nr^\pa|X_\nb^\pa = X_\nr^\pa =
\varnothing,Q)+ \Delta_2 I(X_\nb^\pb;Y_\nr^\pb|X_\na^\pb = X_\nr^\pb =
\varnothing,Q) +\\&\Delta_3
I(X_\na^\pc,X_\nb^\pc;Y_\nr^\pc|X_\nr^\pc = \varnothing,Q)
\end{align*}
over the joint distribution $p(q) p^\pa(x_{\na}|q) p^\pb(x_{\nb}|q)
p^\pc(x_{\na}|q)p^\pc(x_{\nb}|q)$ $p^\pd(x_{\nr}|q)$ over the
alphabet ${\cal X}_\na^2 \times {\cal X}_\nb^2 \times {\cal X}_\nr$
with $|{\cal Q}| \leq 5$. \thmend
\end{theorem}
\myproof{Proof outline:} Generate random codewords ${\bf
x}_\na^{(1)}(w_\na)$, ${\bf x}_\nb^{(2)}(w_\nb)$, ${\bf
x}_\na^{(3)}(w_\na)$, ${\bf x}_\nb^{(3)}(w_\nb)$. Relay $\nr$
receives data from terminal nodes during phases 1 -- 3, which is
decoded by the relay using a MAC protocol to recover $w_\na$,
$w_\nb$. Theorem \ref{theorem:HBC:1} then follows the same argument
as the proof of Theorem \ref{theorem:TDBC:1}.
\endproof
\begin{theorem}\label{theorem:HBC:2}
The capacity region of the bi-directional relay channel with the HBC
protocol is outer bounded by the union of
\begin{align*}
R_\na \leq \min \big\{& \Delta_1 I(X_\na^\pa;Y_\nr^\pa,Y_\nb^\pa|X_\nb^\pa = X_\nr^\pa = \varnothing,Q) + \Delta_3 I(X_\na^\pc;Y_\nr^\pc|X_\nb^\pc,X_\nr^\pc = \varnothing,Q), \\
             &\Delta_1 I(X_\na^\pa;Y_\nb^\pa|X_\nb^\pa = X_\nr^\pa = \varnothing,Q) + \Delta_4 I(X_\nr^\pd;Y_\nb^\pd|X_\na^\pd = X_\nb^\pd = \varnothing,Q)\big\} \\
R_\nb \leq \min \big\{ & \Delta_2 I(X_\nb^\pb;Y_\nr^\pb,Y_\na^\pb|X_\na^\pb = X_\nr^\pb = \varnothing,Q) + \Delta_3 I(X_\nb^\pc;Y_\nr^\pc|X_\na^\pc,X_\nr^\pc = \varnothing,Q),\\
             &\Delta_2 I(X_\nb^\pb;Y_\na^\pb|X_\na^\pb = X_\nr^\pb = \varnothing,Q) + \Delta_4 I(X_\nr^\pd;Y_\na^\pd|X_\na^\pd = X_\nb^\pd = \varnothing,Q)\big\} \\
R_\na+R_\nb \leq &\Delta_1 I(X_\na^\pa;Y_\nr^\pa|X_\nb^\pa = X_\nr^\pa = \varnothing,Q)
        + \Delta_2 I(X_\nb^\pb;Y_\nr^\pb|X_\na^\pb = X_\nr^\pb = \varnothing,Q) +\\&\Delta_3 I(X_\na^\pc,X_\nb^\pc;Y_\nr^\pc|X_\nr^\pc = \varnothing,Q)
\end{align*}
over all joint distributions $p(q) p^\pa(x_{\na}|q)
p^\pb(x_{\nb}|q)p^\pc(x_{\na}, x_{\nb}|q)$ $p^\pd(x_{\nr}|q)$ with
$|{\cal Q}| \leq 5$ over the alphabet ${\cal X}_\na^2 \times {\cal
X}_\nb^2 \times {\cal X}_\nr$. \thmend
\end{theorem}
{\em Remark:} If the relay is not required to decode both messages,
then removing the constraint on the sum-rate $R_\na + R_\nb$ in the
region above yields an outer bound.

\myproof{Proof outline:} The proof of Theorem \ref{theorem:HBC:2}
follows the same argument as the proof of the converse part of
Theorem \ref{theorem:MABC}.
\endproof

\section{The Gaussian Case}
\label{sec:gaussian}

In the following section, we apply the performance bounds derived in
the previous section to the AWGN channel with pass loss. Definitions of codes, rate,
and achievability in the memoryless Gaussian channels are analogous to those of the discrete memoryless
channels. If
$X_\na[k] \neq \varnothing, X_\nb[k] \neq \varnothing, X_\nr[k] =
\varnothing$, then the mathematical channel model is $ Y_\nr[k] =
g_{\na\nr} X_\na[k] + g_{\nb\nr} X_\nb[k] + Z_\nr[k] $ and
$Y_\na[k]$ and $Y_\nb[k]$ are given by similar expression in terms
of $g_{\na\nr}, g_{\nb\nr}$ and $g_{\na\nb}$ if only one node is
silent. If $X_\na[k] = X_\nb[k] = \varnothing$ and $X_\nr[k] \neq
\varnothing$, then  $Y_\na[k] = g_{\nr\na} X_\nr[k] + Z_\na[k]$ and
$Y_\nb[k] = g_{\nr\nb} X_\nr[k] + Z_\nb[k]$ and similar expressions
hold if other pairs of nodes are silent, where the effective complex
channel gain $g_{ij}$ between nodes $i$ and $j$ combines both
quasi-static fading and path loss and the channels are reciprocal,
i.e., $g_{ij} = g_{ji}$. For convenience, we define $G_{ij} :=
|g_{ij}|^2$, i.e. $G_{ij}$ incorporates path loss and fading effects
on received power. Furthermore, we suppose the interesting case that
$G_{\na\nb} \leq G_{\na\nr} \leq G_{\nb\nr}$. Finally, we assume
full Channel State Information (CSI) at all nodes (i.e. each node is
fully aware of $g_{\na\nb}$, $g_{\nb\nr}$ and $g_{\na\nr}$) and that
each node has the same transmit power $P$ for each phase, employs a
complex Gaussian codebook and the noise is of unit power, additive,
white Gaussian, complex and circularly symmetric. For convenience of
analysis, we also define the function $C(x) := \log_2 (1+x)$.

\begin{figure}[t]
\begin{center}
\epsfig{keepaspectratio = true, width = 3.2 in, figure =
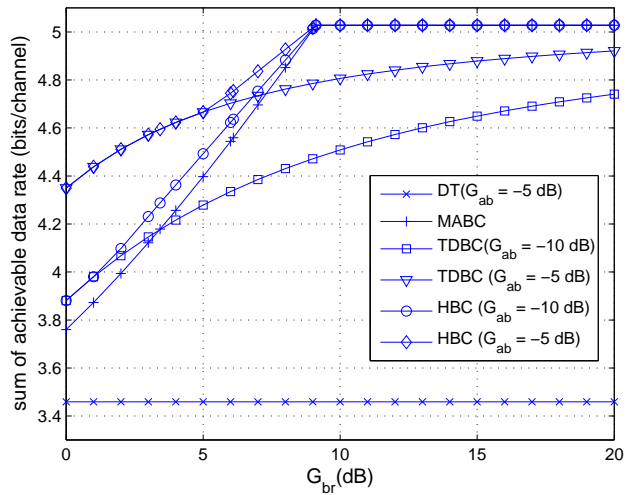}
\end{center}
\caption{Achievable sum rates of the protocols ($P = 15$ dB,
$G_{\na\nr} = 0$ dB)} \label{fig:compare_15db}
\end{figure}

%\subsection{Time optimization}
For a fading AWGN channel, we can optimize the $\Delta_i$'s for
given channel mutual informations in order to maximize the
achievable sum rate ($R_\na + R_\nb$). First, we optimize the time
periods in each protocol and compare the achievable sum rates
obtained to determine an optimal transmission strategy in terms of
sum-rate in a given channel. For example, applying Theorem
\ref{theorem:TDBC:1} to the fading AWGN channel, the optimization
constraints for the TDBC protocol are\footnotemark:
 \begin{align}
 R_\na &\leq \min\left\{\Delta_1 C(PG_{\na\nr}), \Delta_1
 C(PG_{\na\nb}) + \Delta_3 C(PG_{\nb\nr}) \right\}\label{AWGN:4}\\
 R_\nb &\leq \min\left\{\Delta_2 C(PG_{\nb\nr}), \Delta_2
 C(PG_{\na\nb}) + \Delta_3 C(PG_{\na\nr}) \right\}\label{AWGN:5}
 \end{align}
We have taken $|{\cal Q}| = 1$ in the derivation of \eqref{AWGN:4}
and \eqref{AWGN:5}, since a Gaussian distribution simultaneously
maximizes each mutual information term individually as each node is
assumed to transmit with at most power $P$ during each phase. Linear
programming may then be used to find optimal time durations.
\footnotetext{The power constraint is satisfied almost surely as $n
\rightarrow \infty$ in the random coding argument for Gaussian input
distributions with $E[X^2] < P$.} The optimal sum rate corresponding
to the inner bounds of the protocols is plotted in \Fig
\ref{fig:compare_15db}. As expected, the optimal sum rate of the HBC
protocol is always greater than or equal to those of the other
protocols since the MABC and TDBC protocols are special cases of the
HBC protocol. Notably, the sum rate of the HBC protocol is strictly
greater than the other cases in some regimes. This implies that the
HBC protocol does not reduce to either of the MABC or TDBC protocols
in general.

\begin{figure}[t]
\begin{center}
\epsfig{keepaspectratio = true, width = 3.2 in, figure =
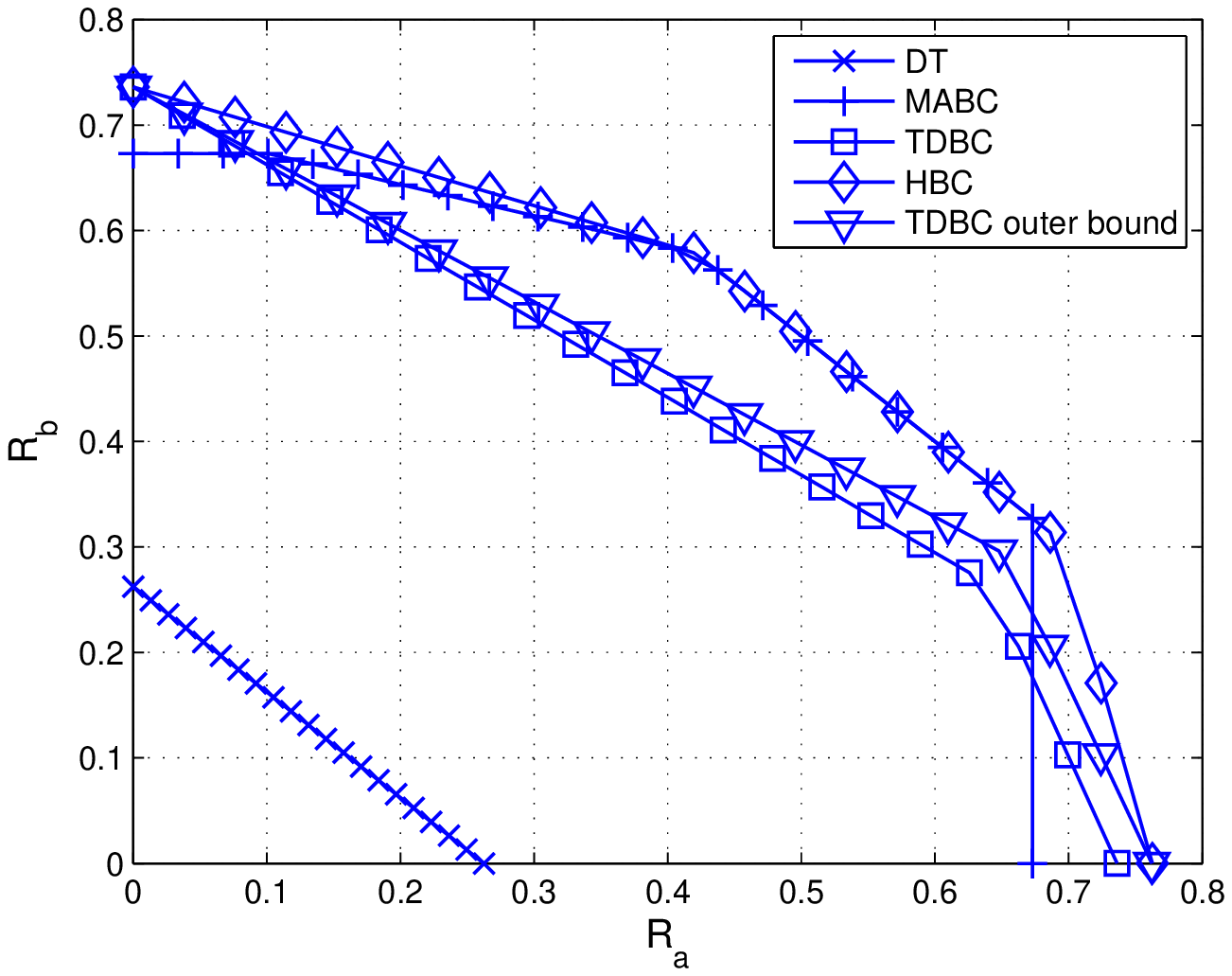}\\
\epsfig{keepaspectratio = true, width = 3.2 in, figure =
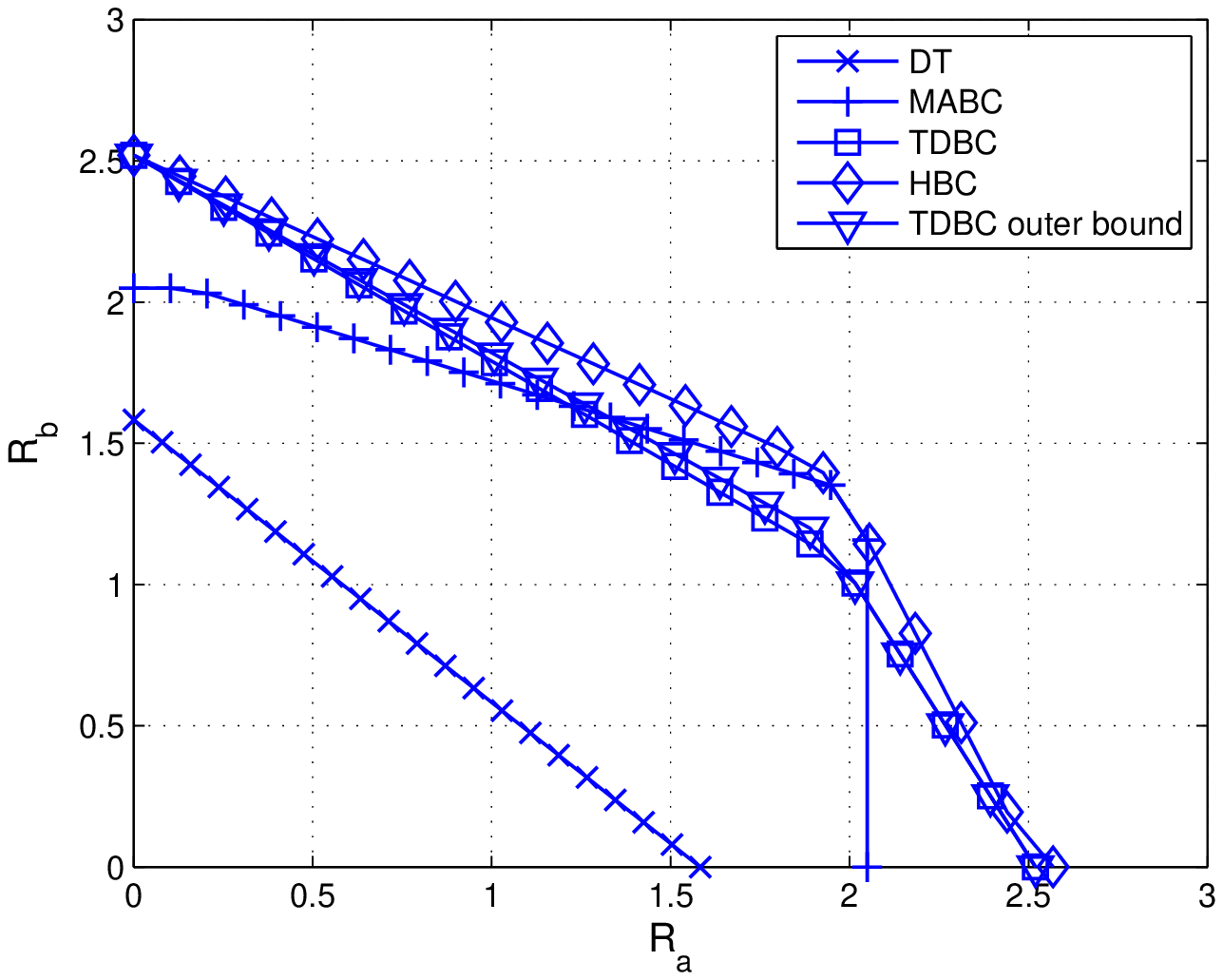}
\end{center}
\caption{Achievable rate regions and outer bounds with $P = 0$ dB
(top) and $P = 10$ dB (bottom) ($G_{\na\nr} = 0$ dB, $G_{\nb\nr} =
5$ dB, $G_{\na\nb} = -7$ dB)} \label{fig:region_1}
\end{figure}

%\subsection{Achievable regions and outer bounds}
In the MABC protocol, the performance region is known. However, in the other
cases, there exists a gap between the expressions. An achievable region
of the 4 protocols and an outer bound for the TDBC
protocol is plotted in \Fig \ref{fig:region_1}
%(in the low SNR regime) and \Fig \ref{fig:region_2}
%(in the high SNR regime).
(in the low and the high SNR regime). As expected, in the low SNR
regime, the MABC protocol dominates the TDBC protocol, while the
latter is better in the high SNR regime. It is difficult to compute
the outer bound of the HBC protocol numerically since, as opposed to
the TDBC case, it is not clear that jointly Gaussian distributions
are optimal due to the joint distribution $p^\pc(x_\na,x_\nb|q)$ as
well as the conditional mutual information terms in Theorem
\ref{theorem:HBC:2}. For this reason, we do not numerically evaluate
the outer bound. Notably, some achievable HBC rate pairs are outside
the outer bounds of the MABC and TDBC protocols.
\bibliographystyle{IEEEtranS}
\bibliography{sang}
\end{document}